\documentclass[aps,prd,preprintnumbers,superscriptaddress,nofootinbib,floatfix]{revtex4-1}

\usepackage[dvipsnames]{xcolor}
\usepackage{amssymb,amsmath,latexsym,mathrsfs}
\usepackage[USenglish,american]{babel}
\usepackage{bm}
\usepackage{amsfonts}
\usepackage{graphicx}
\usepackage{epstopdf}
\usepackage{hyperref}
\usepackage{array}
\usepackage[utf8]{inputenc}
\usepackage{soul}
\usepackage{color}
\usepackage{slashed}
\usepackage{csquotes}
\usepackage[T1]{fontenc}

\enlargethispage{\baselineskip}

\begin{document}

\preprint{NORDITA 2020-088}
\preprint{UTTG-12-2020}

\bigskip
\bigskip
\bigskip
\bigskip
\bigskip

\title{Filling the Black Hole Mass Gap: Avoiding Pair Instability in Massive Stars through Addition of Non-Nuclear Energy}

\author{Joshua Ziegler}
\affiliation{Physics Department, University of Texas, Austin, TX 78712} \email{jjziegler@utexas.edu}
\author{Katherine Freese}
\affiliation{Physics Department, University of Texas, Austin, TX 78712}
\affiliation{The Oskar Klein Centre for Cosmoparticle Physics, Department of Physics, Stockholm University, SE-106 91 Stockholm, Sweden}
\affiliation{The Nordic Institute for Theoretical Physics (NORDITA) Roslagstullsbacken 23, SE-106 91, Stockholm, Sweden}
\email{ktfreese@utexas.edu}

\begin{abstract}
In standard stellar evolution, stars with masses ranging from approximately 150 to $240 M_\odot$ are expected to evolve to a pair instability supernova with no black hole (BH) remnant. This evolutionary behavior leads to a predicted gap in the black hole mass function from approximately 50 to $140 M_\odot$. Yet the LIGO and Virgo Collaborations\cite{ligo2020} recently discovered black holes of masses $66 M_\odot$ and $85 M_\odot$ in the gravitational wave event GW190521.  We propose a new method to populate the BH mass gap. If an energy source is added throughout the star in addition to nuclear fusion, it is possible for the altered evolution to avoid the complete destruction of a pair instability supernova, and instead a BH remnant is left behind.
An example of an extra energy source is dark matter annihilation within the star, but our results hold more generally. We show this phenomenon by exploring the effect of adding an energy source independent of temperature and density to a $180 M_\odot$ star, using the MESA one-dimensional stellar evolution software. If $\sim$50\% of the star's energy  is due to this new source, the star is capable of avoiding the pair instability entirely and evolving towards a core-collapse supernova and ultimately a BH remnant with mass $\sim 120 M_\odot$.
\end{abstract}

\maketitle
	
\section{Introduction}

The discovery of gravitational waves (GW) from black hole (BH) 
mergers by LIGO in 2015\cite{ligo2015} has opened a new window into the Universe. The subsequent GW events found by the  LIGO and Virgo Collaborations are teaching us more and more about the stellar-mass black hole mass function, i.e. 
the number of black holes that exist as a function of BH mass. 
Theoretical predictions for the precise details of the mass function would require a thorough accounting of the formation rate of stars, their evolution into black holes, and the merger history of those black holes. 

In this paper we focus on the evolution of individual stars into black holes, with the twist that we add an additional energy source as well as standard fusion. 
Previous authors \cite{woosley2002}\cite{woosley2014, woosley2017, woosley2019, belczynski2016, spera2017} studied the evolution of standard fusion-powered stars into black hole remnants  and found that there should be no black holes in the so-called ``upper mass gap'' due to pair instability supernovae. 
In stars with a main-sequence mass of approximately\footnote{For concreteness, we will use the boundaries reported in \cite{woosley2014}, but these are subject to an uncertainty of up to approximately $10 M_\odot$ depending on the particular method used when performing numerical simulations.} 150-$260 M_\odot$, the conversion of photons to $e^+/e^-$ pairs inside the hot dense core drives a runaway collapse. When this collapse is halted by oxygen-burning nuclear reactions, the energy produced leads to an explosion powerful enough to completely destroy the star, leaving no remnant. Thus, standard stellar astrophysics tells us there should be no black holes with masses in the range 50-$140 M_\odot$. The recent discovery of GW190521 by the LIGO and Virgo Collaborations\cite{ligo2020, ligo2020a} of black holes with masses $66 M_\odot$ and $85 M_\odot$ raises questions about this model. 
It should be noted that the GW190521 data is consistent with a model in which one of the black holes is below the mass gap and the other above\cite{fishbach2020}; 
however, preference for this model over the LIGO/Virgo reported masses requires the addition of a prior
 assumption that at least one of the black holes is not in the mass gap. Furthermore, by extending the analysis done by the LIGO/Virgo collaborations to allow for high-eccentricity merger scenarios, masses of approximately $120 M_\odot$, which are still within the mass gap, are preferred \cite{gayathri2020}.

Overcoming the destructiveness of pair instability supernovae to populate this upper mass gap with black holes proves a challenge. Although it is possible to move the boundaries of the mass gap by including rotation or magnetic fields in the calculation\cite{branch}, physically reasonable values of these typically cannot allow isolated stars to populate the full gap. As a result, even taking into account the uncertainty in the mass gap's edges\cite{farrell2020}, observing a black hole with a mass of approximately $100 M_\odot$ would still be unexpected from stellar evolution.

Here we propose a new  mechanism to populate the pair instability mass gap. We consider the effect of adding a new source of energy to the star, in addition to nuclear fusion.  This non-nuclear energy prevents the star from completely blasting apart as a pair instability supernova and leaves behind a BH remnant instead.
We are motivated to introduce this energy source by the dark matter heating that can exist in the earliest stars, but our analysis is general and energy sources with entirely different origins could show similar behavior. 
With the addition of a new energy source, a star that would explode as a pair instability supernova according to standard stellar evolution, would instead evolve into a core-collapse supernova and ultimately leave behind a black hole or (rarely, due to the masses involved) neutron star remnant.
Below in section~\ref{results} we specifically study the case of a $180 M_\odot$ star, which in standard fusion-powered stellar evolution,  ends its life in a pair instability supernova that completely destroys the star.
We ran the MESA (Modules for Experiments in Stellar Astrophysics) stellar evolution code to study stars of this mass, both with and without the addition of a non-nuclear power source.
We show that when $\sim 50\%$ of the stellar luminosity arises from the new heat source, the star is no longer completely destroyed.  
With  our modified stellar evolution code, we were able to follow the evolution of a star with initial mass $180M_\odot$ to a core collapse supernova precursor  with mass $119 M_\odot$,
which is expected to leave behind a black hole remnant with a slightly lower mass. As a result, we find that a star with extra non-nuclear energy may evolve into a black hole in the mass gap. 

Others have proposed alternative mechanisms, going beyond the collapse of a single isolated star,  to explain BH in the mass gap.
One approach is mergers of black holes \cite{fragione2020} or their stellar precursors \cite{carlo2019, carlo2019short, vanSon2020}.
Primordial black holes (PBH) in this mass range were predicted by \cite{carr2019}; or PBHs could merge to produce the observed BH \cite{luca2020}.
Further, extensions of the Standard Model were proposed as another explanation \cite{sakstein2020}.

An outline of the remainder of this paper is as follows. In section \ref{stars}, we describe in greater detail the standard evolution of stars without the non-nuclear energy source that we consider. We then describe in section \ref{dark stars} the phenomenon of dark stars which we use as a motivating case for our study. We follow this in section \ref{mesa} with a description of our numerical methods. Finally, in section \ref{results}, we provide comparisons between the stars that include and do not include the non-nuclear energy source.

\section{Standard Evolution of Massive Stars Powered only by Fusion}\label{stars}

In this section, we review   the standard evolution of massive fusion-powered stars. (In later sections, we will modify the standard picture by adding an additional heat source.)
We examine stars with a total mass greater than $80 M_\odot$, as these are the stars whose evolution will be impacted by pair instability, a phenomenon that we discuss in section II.C.
Of particular interest are the stars with a total mass of  $\sim 150 - 240  M_\odot$, as these are the stars that blast apart as pair instability supernovae and lead to the BH mass gap. 
With that in mind, we focus on stars that have metallicity $Z=0$, since stars with even a relatively small fraction of solar metallicity rarely reach such large masses\cite{woosley2002}. In addition, for simplicity, the stars we look at will be non-rotating and we will ignore magnetic effects. While adding in rotation or magnetic fields may change the masses of the boundaries of the mass gap, realistic treatments do not affect the stars enough to change our general results. With these assumptions, we can treat stars as spherically symmetric, and therefore one-dimensional.

The evolutionary behavior of such stars can be described in three main stages: a protostellar phase, a quasi-stable phase, and at least one period of collapse and subsequent explosion. In some stars, this last phase is caused by pair instability, while in others it is caused by fusion involving iron nuclei. We provide here a brief overview of each of these phases in completely nuclear-powered stars\cite{kippenhahn, hansen, schwarzschild, phillips}.

\subsection{Protostellar Evolution}

Before reaching a stable state, the protostellar gas cloud that will eventually form a star gradually collapses under the force of its own self-gravity. In the process, the temperature and density at the center of the gas cloud increase. As the density increases, nuclei collide more frequently, and as the temperature increases, the collisions become more energetic. Both of these trends increase the rate at which the exothermic nuclear reactions converting hydrogen into helium occur, while simultaneously reducing the reverse reactions. As a result of the increased energy output of these processes and the steepening temperature and density profiles, pressure within the star becomes strong enough to counteract the force of gravity, and the gas cloud becomes a stable star.

\subsection{Quasi-Stable Evolution}

Once stable, a star spends most of its lifetime in approximate hydrostatic and thermal equilibrium, and these conditions can be used to determine the stellar evolution. In particular, by using the equilibrium conditions, it is possible to determine the radial profiles of pressure $P(r)$, temperature $T(r)$, luminosity $L(r)$, and mass $M(r)$ contained within a sphere of radius $r$ at a given instant of time. As it turns out, calculating these profiles is often more practical if the dependent variable is not the radius but the mass contained within that radius. As a result, we can gather profiles of pressure, temperature, luminosity, and radius as a function of mass at any given time using a set of coupled PDEs that may be written \cite{kippenhahn}
\begin{align}
	\frac{\partial r}{\partial M} &= \frac{1}{4 \pi r^2 \rho}, \label{rm}\\
	\frac{\partial P}{\partial M} &= - \frac{GM}{4\pi r^4}, \label{Pm}\\
	\frac{\partial L}{\partial M} &= \epsilon, \label{Lm}\\
	\frac{\partial T}{\partial M} &= -\frac{GMT}{4 \pi r^4 P} \nabla.\label{Tm}
\end{align}

In addition to the four functions that we are solving for, an additional three functions must be introduced to close the system of equations: $\rho(P,T,x_i), \epsilon(\rho, T, x_i),$ and $\nabla(\rho, T, x_i)$, where $x_i$ describes the abundances of each nuclear isotope in the star. The function $\rho$ is determined from the equation of state for the gases which make up a star in the range of temperatures, densities, and chemical abundances that the star may experience. In most cases, it is appropriate to use as the equation of state a sum of the equations of state of various gases: an ideal gas to describe ionized nuclei and atoms, a Fermi-Dirac gas to describe any free electrons and positrons in the star, and a photon gas to describe the radiation throughout the star\cite{fowler1964, blinnikov1996}. 

The function $\epsilon$ describes the amount of energy generated per unit time per unit mass within the star (in cgs units, this function will have units of erg/g/sec). For our present purposes, this function includes nuclear energy production and any energy lost to neutrinos, as well as the energy generated by gravitational changes. We will also add an additional constant energy source below, though we do not include it for this discussion of stars powered only by nuclear energy. Of particular interest, the nuclear energy production rate depends strongly on the temperature and density, so the radial profile of the energy production rate is strongly peaked near the high-temperature center of the star\cite{fowler1964}.

Finally, the function $\nabla$ encodes the transport of energy through the star. Depending on whether the predominant method of transport is convection, radiation, or conduction, $\nabla$ can take different forms\cite{kippenhahn}.
\begin{equation}
	\nabla = \begin{cases}
		\frac{\partial \ln T}{\partial \ln P} & \mathrm{convection}\\
		\frac{3}{16\pi a c G}\frac{\kappa L P}{M T^4} & \mathrm{radiation, conduction}
		\end{cases}.
\end{equation}
Here, the opacity $\kappa$ is a function of temperature, density, and chemical composition. To determine which case to use, a prescription for determining what kind of heat transfer dominates is necessary. We use the Ledoux criterion for dynamic stability for this purpose\cite{ledoux1947}.

Implicit throughout all of these functions has been a dependence on the chemical composition, described by the isotope abundances, $x_i$. These abundances themselves vary throughout the star. It is only through changes in this chemical composition that evolution can occur while a star is in this quasi-stable phase. As nucleosynthesis changes the composition throughout the star, the profiles of temperature, pressure, radius, and luminosity must change to compensate. 

During this phase of the evolution, the stellar structure typically changes gradually, but as the star evolves, it undergoes periods of rapid contraction and expansion. During these periods, the hydrostatic equations aren't enough to describe the evolution. To remedy this, equation \ref{Pm} can be replaced by 
\begin{equation}
\frac{\partial P}{\partial M} = - \frac{GM}{4\pi r^4} - \frac{\partial^2r}{\partial t^2} \frac{1}{4\pi r ^2}. 
\end{equation}
At the same time, the function $\epsilon$ must be expanded to include additional terms that reflect the time-dependence of temperature and pressure ($\propto \frac{\partial T}{\partial t}$ and $\propto \frac{\partial P}{\partial t}$ respectively). These terms are necessary to account for the non-adiabatic nature of a collapse or explosion\cite{kippenhahn}.

\subsection{Pair Instability}

In normal stars, collapses and explosions typically occur when changes in chemical composition prevent nuclear reactions from providing enough pressure to fully support the star.
In the range of masses we study, stars may also experience collapse and explosion due to changes in the equation of state brought about by production of electron-positron pairs\cite{rakavy1967}. Pressure support within the star can be attributed to the combined effects of radiation pressure and gas pressure. The question of whether collapse will occur is closely related to the adiabatic coefficient $\gamma = \partial\ln P / \partial\ln \rho$. Because both pressure $P(M)$ and density $\rho(M)$ are functions of the mass $M$ enclosed within a spherical shell, $\gamma(M)$ also varies throughout the star. As a result, if any mass shell $M_{<4/3}$ satisfies $\gamma(M_{<4/3}) <4/3$ that mass shell will be unstable and will undergo hydrodynamic collapse (as opposed to hydrostatic contraction). We identify the instability that causes a mass shell to collapse as a ``pair instability.''

\begin{figure}
	
	\includegraphics[width=\textwidth]{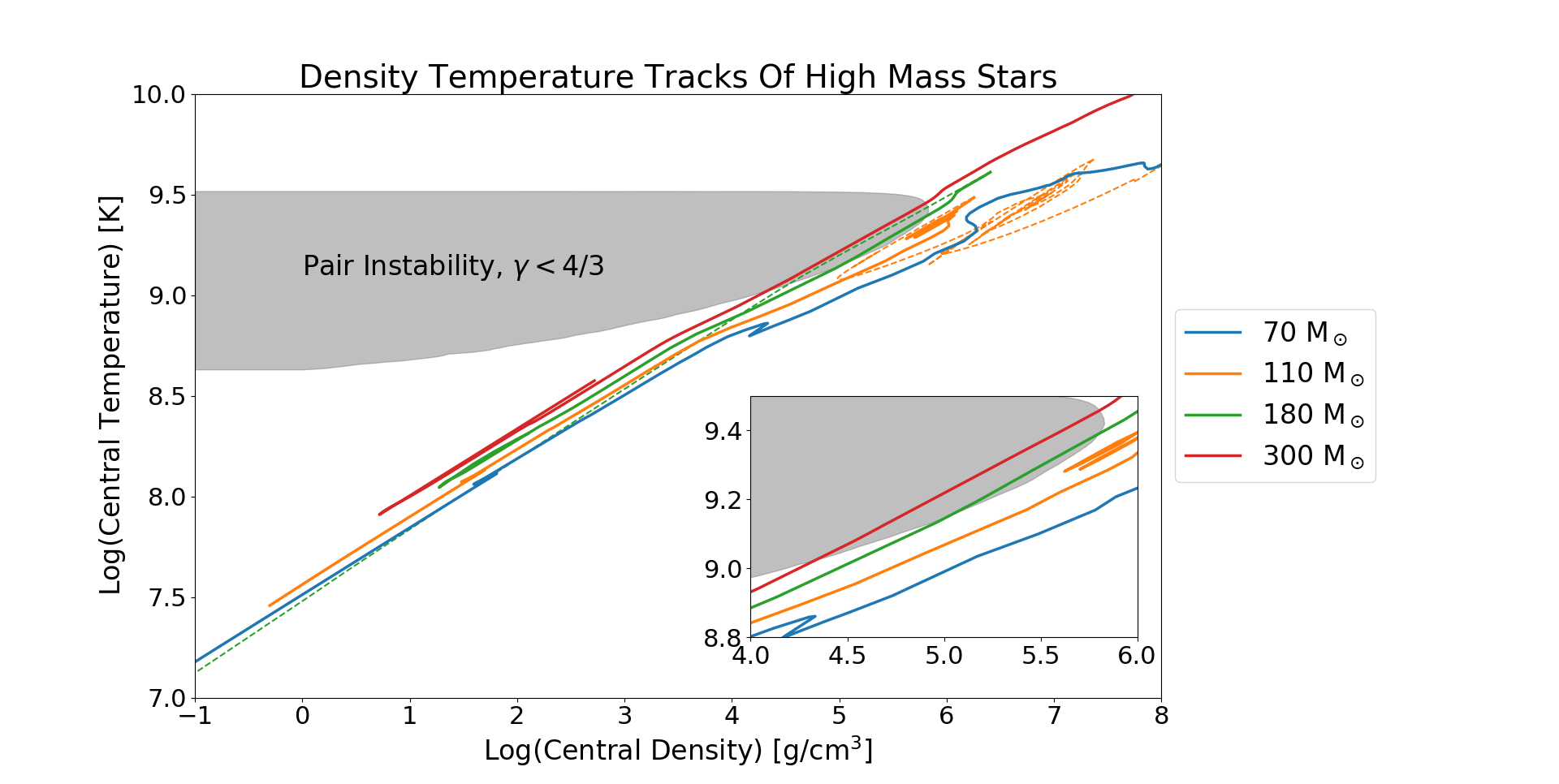}
	
\caption{
\label{trho}
Standard stellar evolution for stars of different masses powered only by nuclear fusion. Each solid curve follows the evolution of the central density and central temperature from low density/low temperature in the lower left to higher density and higher temperature in the upper right. During their evolutions, all of these stars approach the conditions necessary for pair instability to occur somewhere within the core of the star. Unlike the other three stars, the $70 M_\odot$ star (green curve) approaches, but does not satisfy, those conditions. As a result, its evolution is not affected by the pair instability, and it evolves to a core collapse supernova, leaving behind a BH remnant. The other three stars do reach the temperature and density conditions necessary to undergo the pair instability somewhere in the core. In the $110 M_\odot$ star (orange curve), these conditions are met only in mass shells outside the center of the star. As a result, only the outermost layers of the star are affected by the pair instability, and the star undergoes a pulsational pair instability supernova (PPISN). In this type of supernova, large ejections of mass occur due to explosions induced by the pair instability, but a portion of the mass of the star remains and evolves to a core-collapse supernova, which leaves behind a BH remnant. In the $180 M_\odot$ star (green curve), the entire core of the star is affected by the pair instability, including the center of the star. As a result, the star undergoes a pair instability supernova (PISN), in which the entire star explodes, and there is no BH remnant. The $180 M_\odot$ star is the only case shown where no BH remnant is left behind. In the $300 M_\odot$ star (red curve), the pair instability also affects the entire star, but before an explosion ends the pair instability collapse, the core begins collapsing toward a core-collapse supernova. As a result, this star collapses leaves behind a BH. In addition to the evolutionary tracks, this figure highlights the pair instability region (gray shaded region), in which $\gamma < 4/3$. Evolutionary tracks which pass through this region in the figure (180 and $300 M_\odot$) undergo pair instability throughout the core, including at the center of the star. Stars that pass near this region but do not enter it ($110 M_\odot$) may still undergo pair instability, but only in mass shells outside the center of the star. And stars that pass farther from this region ($70 M_\odot$) do not experience pair instability anywhere in the star. The inset figure shows the evolutionary tracks of each star the first time they pass through or around the pair instability region. For stars that underwent an explosion due to the pair instability, their evolution after the onset of the (first) explosion is shown as a dashed line. Ultimately, every star except the $180 M_\odot$ star finished its evolution in the upper right of the plot, corresponding to a core-collapse supernova and a BH remnant. The $180 M_\odot$ star, on the other hand, finished its evolution back in the lower left corner of the plot, indicative of a pair instability supernova and no BH remnant. 
}
	
\end{figure}

As a star evolves, electron-positron pairs are produced in any mass shell within the star where local temperature and density conditions allow it. The formation of electron-positron pairs can lower the total pressure support within the star. If enough electron-positron pairs are produced in a mass shell, then, the drop in pressure support can become significant enough for that mass shell to become unstable (i.e. $\gamma <4/3$) and to begin to collapse. Because there are physical restrictions on what temperature and density conditions may give rise to electron-positron pairs in a star, only certain combinations of density and temperature will lead to a collapsing mass shell (due to pair production). The collection of densities and temperatures which lead to collapse form a closed region in density-temperature space, which can be seen as the gray shaded region in the upper left portion of Figures~\ref{trho} and \ref{trhoex}.  We identify this region in density-temperature space, which is defined by the equation of state within the star and does not depend on where in a star we look, as the ``pair instability region.''

The boundaries of this region arise in two ways. First, in order for the reaction $\gamma \gamma \rightarrow e^+ e^-$ to occur, the energy of the photons must be higher than the combined rest mass of the electron and positron. In low-temperature regions, this condition is met only in the exponentially suppressed Boltzmann tail of the photon energy distribution, effectively preventing the reaction from occurring. As a result, there is a lower limit to the temperature at which electron-positron pairs are produced. The rest of the shape of this shaded region arises largely because of Fermi blocking of the electrons. Because electrons are fermions, the available end states of the reaction can be reduced by Pauli exclusion. As a result, in high density regions, where the ionized electron density is also high, higher temperatures are necessary to encourage the same reaction rate of photons to electron-positron pairs. In a similar way, at higher temperatures, existing electron-positron pairs can partially block the formation of additional pairs, so that the relative energy stored in photons increases with increasing temperature. In other words, even though electron-positron pairs continue to be produced as the temperature increases, the photon abundance increases more quickly, so that the region of the star under consideration feels the effects of the pair production less strongly. This ultimately leads to the upper boundary on the shaded region.

``Pair instability'' and ``pair instability region'' are based on local temperature and density conditions within a mass shell, but what we are most interested in is evolutionary behavior affected by the collapse of mass shells due to pair instability. To determine whether pair instability will lead to any noticeable evolutionary changes, we need to consider how much of a star is affected by pair instability. If very little of the star undergoes pair instability, the evolutionary effect will largely be negligible, and the star will evolve like stars in which no pair production occurred at all. On the other hand, if much of the star undergoes pair instability, the evolutionary behavior can be drastically different than that of a star in which no pair production occurs.

Pair instability affects stars with different masses in different ways, leading to multiple distinguishable evolutionary behaviors. As expanded upon below, possible evolutionary endstates of stars that undergo pair instability include core collapse supernovae,  pulsational pair instability supernovae (PPISNe), and pair instability supernovae (PISNe). PPISNe are characterized by explosive nuclear burning leading to an incomplete explosion of the star. What is left behind, however, does not collapse immediately into a black hole, but re-evolves as a smaller mass star. As a result, PPISNe can involve multiple periods of pair instability, ejecting a portion of the star's mass each time. Ultimately, though, PPISNe typically evolve to a core collapse supernova, leaving behind a black hole.  While core collapse supernovae and PPISNe typically leave behind a collapsed remnant, either a neutron star or black hole, PISNe leave no remnant.

\subsection{Evolution Off of the Main Sequence}

For much of their lifetimes, stars burn hydrogen into helium on the main sequence, regardless of their mass, but as stars get older and begin to evolve away from the main sequence, their evolutionary behavior begins to diverge, ultimately leading to quite different endstates. One possible evolution concludes with the star collapsing due to runaway fusion reactions in the iron core. This collapse triggers an explosion known as a core-collapse supernova. This type of supernova typically ejects some mass from the star, but still leaves behind a black hole or neutron star. Another possible outcome of stellar evolution is a collapse due to the pair instability. This type of collapse also triggers fusion reactions in the core, but in these collapses, the core is primarily composed of carbon and oxygen. As a result of this difference in composition the collapse induced by pair instability leads to an explosion that ejects at least some of the mass of the star. If the explosion is powerful enough, it prevents the formation of a black hole remnant. The primary factor in determining what evolutionary behavior a star will follow is mass, though factors like rotation and magnetic field strength can also play a role. Below, we describe the evolution that leads to each type of endstate according to the main sequence stellar mass. 

\subsubsection{Stars with masses below $80 M_\odot$: Core Collapse}

Stars with masses below $\sim80 M_\odot$ eventually evolve to core collapse supernovae, leaving behind black holes or (for lower mass stars) neutron stars. Although stars near $80 M_\odot$ may achieve the conditions necessary for electron-positron pairs to form in small regions throughout the star, the small scale on which these conditions occur typically do not allow any kind of large-scale evolutionary effect. As a result, the evolution they experience once they leave the main sequence is fairly straightforward. As fuel in the core is depleted through nuclear fusion, the chance of nuclei reacting and the energy produced by those reactions decrease. As the evolution continues, the star contracts, increasing its temperature and density. If the adiabatic temperature and density increase is not enough to compensate for the lower amount of nuclear material, the core of the star collapses non-adiabatically. As the temperature and density increase, additional nuclear reactions become accessible and begin to occur rapidly. This cycle occurs iteratively as each nuclear fuel is depleted, causing the central cores of these stars to gradually become dominated by heavier elements.

Once the core of such a star has developed to include a non-negligible amount of Fe-56, this repetitive process of building up heavier nuclei in the core ends. Fe-56 cannot be used as fuel in any exothermic reactions, but can be fuel for endothermic reactions. As a result, once the temperature and density have become high enough that nuclear reactions with iron as a fuel become non-negligible, energy is lost from the radiation in the core. However, this loss of radiation leads to a reduction in pressure support within the star, and the star begins to collapse. As it collapses, the temperature and density further increase, causing the endothermic nuclear reactions to occur yet more rapidly. This cycle leads to a runaway collapse that ultimately leads to a core-collapse supernova. In general, core collapse supernovae may evolve to either neutron stars or black holes, but given the stellar mass necessary for pair instability to be of interest, the core collapse supernovae we look at typically evolve to black holes.

\subsubsection{Stars with masses $80 M_\odot$ to $150 M_\odot$: Pulsational Pair Instability (PPISN)}

Stars with masses between $80 M_\odot$ and $150 M_\odot$, after a series of explosions known as a Pulsational Pair Instability Supernova (PPISN), also evolve to core collapse supernovae and leave behind black holes (and, in an extremely narrow mass range, neutron stars). The larger number of electron-positron pairs produced in the star, compared to lower mass stars, causes the star to show evolutionary changes due to the pair instability. In particular, pair instability in mass shells outside the center of the star causes a collapse of the outer regions of the core. This collapse triggers an increase in oxygen-burning nuclear reactions in the outer portion of the core. The energy from these oxygen reactions causes the star to become partially unbound, with outer layers being ejected but the inner portion of the core remaining bound.
The material that is ejected is predominantly composed of hydrogen and helium from the stellar envelope and outer layers of the helium shell, and is believed to be ejected nearly spherically symmetrically.

What remains of the star gradually recollapses into a stable state. This new, lower mass star evolves as before, gradually fusing elements in the core until mass shells within the core approach pair instability again. Depending on this new star's mass, it may repeat this cycle of pair-instability-induced collapse, ejection, and recollapse multiple times, each time shrinking the remaining mass of the star. Eventually, this process leads the star to have a low enough mass that it completely avoids the evolutionary effects of the pair instability. Once the star is low enough mass to avoid the pair instability everywhere, it continues fusing elements in the core, until a sizable iron core has developed, and then undergoes core collapse. This time, the iron-fusion reactions cause the star to undergo a core collapse supernova, which leaves behind a black hole, or very rarely a neutron star. However, because of the mass loss that has happened due to the pair instability induced ejections, the mass of a black hole produced by this pulsational pair instability process will not have a mass greater than $\sim 50 M_\odot$.

\subsubsection{Stars with masses $150 M_\odot$ to $240 M_\odot$:  Pair Instability Supernova (PISN) and no Black Hole Remnant}

For more massive stars, the pair instability can lead to a full destabilization of the star. Like stars that go through a PPISN, stars in the mass range $150 M_\odot$ to $240 M_\odot$ undergo a pair instability. Unlike stars that go through PPISN, the pair instability affects a larger portion of these stars' cores, typically including the center of the star. In addition, the collapse begins slightly earlier, so less carbon has been depleted. Both factors lead to a more powerful explosion once oxygen burning halts the collapse. In fact, the explosion is powerful enough to cause the entire star to have a velocity greater than the local escape velocity and to be completely destroyed. If observed, this explosion would be a pair instability supernova (PISN). As the entire mass of the star explodes outward, the approximately spherical shell of material will be dense with nuclear composition ranging from hydrogen to nickel. A PISN leaves behind no black hole remnant.

\subsubsection{Stars with masses above $240 M_\odot$: Core Collapse}

For even larger stars, the combined effects of pair instability and iron fusion lead the star to evolve to a core collapse supernova and to leave behind a black hole remnant. As the star collapses due to undergoing the pair instability, the energy released by nuclear reactions competes with this collapse. However, before the nuclear reactions can cause the collapse to halt, iron is accumulated in the core. As the collapse continues, the core of the star reaches temperatures and densities high enough to allow for endothermic nuclear processes involving that iron, even while regions outside of this inner portion of the core are involved in exothermic processes. The runaway core collapse due to the iron reactions win out, and the star collapses toward a core collapse supernova.  Remnant black holes produced in this way have masses approximately $140 M_\odot$ or greater.

\section{Dark Stars: A Case Study in non-Nuclear Energy Production}\label{dark stars}

The aim of this paper is to examine the consequences of significant energy injection into stars, beyond that of standard fusion processes. As an example, a specific model with an extra energy source is the energy produced by dark matter annihilation inside early stars, the mechanism by which Dark Stars can exist.  However, many of the results of our paper should generalize to any stellar energy source powerful enough to provide $\sim $ half the energy of the star, so long as that energy production mechanism is not strongly dependent on temperature and density.

Dark Stars are stellar objects which form early in the history of the Universe (typically z=10-50) during the evolution of protostellar clouds in the dark-matter-rich centers of $10^6-10^8 M_\odot$ minihaloes \cite{spolyar2008, freese2008, spolyar2009, freese2010}
(for a review, see \cite{freese2016}). As gas collapses in this protostellar phase, the changing gravitational potential well can cause dark matter to be drawn inward in a process that can be described most simply according to adiabatic contraction\cite{blumenthal1986}. The result of this process is ultimately a density of dark matter throughout the protostellar gas cloud that is higher than the background galactic distribution would predict. In several well-motivated models of dark matter, including Weakly Interacting Massive Particles (WIMPs) and Self-Interacting Dark Matter (SIDM), a natural product of this high density of dark matter is self-annihilation into standard model particles. If the density of protostellar gas is high enough, the annihilation products can become trapped, thermalizing with the gas and providing a substantial source of heat. Due to this heat source, the gas can reach thermal and hydrostatic equilibrium, forming a dark star. 

Although dark matter annihilation can heat a cloud of gas in a similar way as nuclear fusion, the resulting star will have vastly different properties \cite{freese2016, spolyar2009}. Like early nuclear-powered stars, dark stars are almost entirely made of hydrogen and helium gas. However, because dark matter annihilation does not depend on temperature or density nearly as strongly as does nuclear fusion, the dark star becomes stable much earlier in its collapse. This means that dark stars are less dense and cooler throughout than stars supported by fusion. Furthermore, dark stars have a larger radius, $R_{DS} \sim $ 10 A.U. Being cooler throughout, the surface temperature is low, $\sim 10^4$K, too cool to produce significant amounts of ionizing radiation, but because of the large radius the star is very luminous. Without ionizing radiation countering their growth, dark stars may grow from small initial masses, $\sim 1 M_\odot$, to quite large masses, potentially reaching $10^7 M_\odot$ and $10^{10} L_\odot$.

All of these differences between Dark Stars and ordinary fusion-powered stars are related to a fundamental difference in the spacial distribution of energy production within the star. 
Fusion-powered stars produce energy predominantly in a small region in the center of the star due to the strong dependence of the nuclear reaction rate on temperature and density. Compared to this, the energy production in a dark star is spread throughout the star fairly evenly because the energy production is much less dependent on temperature and density. Specifically, for a dark matter model consisting of a dark matter particle of mass $m$, average cross section $\langle\sigma v\rangle$, and self-annihilation branching ratio to neutrinos of $BR_\nu$, the energy production rate as a function of the mass enclosed within a radius ($M$) is\cite{spolyar2009, freese2016}
\begin{equation}
	\epsilon_{DM}(M) = \frac{\partial L}{\partial M} = (1-BR_\nu) <\sigma v> \frac{{\rho_{DM}(M)}^2}{m \rho(M)}.\label{epsdm}
\end{equation}
We subtract off the branching ratio to neutrinos, which describes the fraction of dark matter annihilations that produce neutrino/anti-neutrino pairs, because neutrinos can free stream from the star in all except the most extreme circumstances. As a result, any energy that goes into producing neutrinos is unavailable to be thermalized with the gas in the star.
Here, $\rho(M)$ is the total mass density of the dark star at a radius where the mass enclosed is $M$ and $\rho_{DM}(M)$ is density of only the dark matter at that same location.  

The dark matter density in the center of a protogalaxy can be shown to be approximately\cite{spolyar2009}
\begin{equation}
	\rho_{DM} \approx 5 (\mathrm{GeV/cm}^3)(n / \mathrm{cm}^{-3})^{0.81}
	\label{rhodm}
\end{equation}
where $n$ represents the number density of nuclei. We can relate the $n$ to the mass density $\rho$ by $n \propto \rho/\langle A\rangle$, where $\langle A \rangle$ is the average mass number of the nuclei. Using this to relate equations~\ref{epsdm} and \ref{rhodm}, we find the relation $\epsilon_{DM} (M) \sim \rho^{0.62}/ \langle A \rangle^{1.62}$. 

For comparison, a simple model of nuclear fusion (pp-chain) is $\epsilon_{pp} \propto \rho T^4$\cite{hansen}. Other reactions can be modeled using higher exponents of $T$ and/or $\rho$. Because both $\rho$ and $T$ are decreasing functions of $M$, the nuclear fusion energy production is far more centrally peaked than is the dark matter energy production. 

For simplicity, we used an energy production model $\epsilon_{non-nuc} \propto \mathrm{const}$ in Eqn.(3) for our simulations. While this is not a perfect match to the energy production from dark matter (DM) annihilation,  the results from the two models should be similar.  As discussed above, the energy production in both cases is distributed far more evenly throughout the star than that from nuclear fusion. Fusion only takes place in the hot dense core of the star, whereas DM annihilation is independent of temperature and takes place throughout the star.
In addition, for the stars we consider, the functions $\langle A \rangle$ and $\rho$ are both monotonically decreasing as functions of $M$, so they partially counter one another, leading to a more even overall distribution. 
A similar effect should be seen when adding any energy source that is at most weakly dependent on temperature and density to a nuclear powered star, including the limiting case of complete independence, which is the energy source we used.

\section{Numerical Recipe}\label{mesa}

To perform all of our stellar evolution modeling, both with and without extra non-nuclear energy, we used the one dimensional stellar evolution program MESA (Modules for Experiments in Stellar Astrophysics) version r12778\cite{Paxton2011, Paxton2013, Paxton2015, Paxton2018, Paxton2019}. To generate each main-sequence stellar model, we begin with the model of a known stellar structure, typically a $30 M_\odot$ star with zero metallicity. Mass is gradually added to this seed model until the target mass is reached. In the process of adding this mass, the structure of the star is adjusted to obey the physics that we set, e.g. how much extra energy is being added. Once the target mass is reached, the composition of the model is incrementally adjusted to realign with the desired initial composition.

Although each of the simulations we performed, both with and without the non-nuclear energy source, began with the same $30 M_\odot$ seed star, the main-sequence stars that were built up are all subtly different, even when comparing main-sequence stars with the same mass. The primary reason for these differences is that the process by which the $30 M_\odot$ seed stars are grown to their target mass incorporates the physics that will be used to evolve the stars. For our purposes, this means that as mass is added to the seed star, at each step, the star must satisfy the stellar stability equations (Equations~\ref{rm}-\ref{Tm}) with different amounts of non-nuclear energy added. Although this behavior is not necessarily ideal, it usually ensures that the main sequence star generated by this process can be evolved forward in time without immediately becoming numerically unstable. 

From this initial main sequence model, the star can be evolved step by step. At each timestep, the structure of the star is calculated using an equation of state blended from the OPAL \cite{Rogers2002}, SCVH \cite{Saumon1995}, PTEH \cite{Pols1995}, HELM \cite{Timmes2000}, and PC \cite{Potekhin2010} equations of state. Radiative opacities are primarily derived from OPAL data \cite{Iglesias1993, Iglesias1996}, with low-temperature data from \cite{Ferguson2005} and high-temperature data in the Compton-scattering dominated regime by \cite{Buchler1976}. Meanwhile, electron conduction opacities are from \cite{Cassisi2007}. 

Computing the energy production rate from nuclear reactions is a computationally expensive task, so it is simplified by considering only some of the nuclear isotopes. Within MESA, these nuclear reactions are treated using tabulated values from NACRE \cite{Angulo1999}, JINA REACLIB \cite{Cyburt2010}, plus additional tabulated weak reaction rates \cite{Fuller1985, Oda1994, Langanke2000}. Specifically, we use the ``approx21'' nuclear network containing 21 isotopes with 116 reactions between them, optimized to ensure energy production rates are accurate. We include screening under the prescription of Chugunov\cite{Chugunov2007} and thermal neutrino loss from \cite{Itoh1996}.
The extra energy we introduce is incorporated simply by adding a constant term to the right-hand side of equation~\ref{Lm}. In doing this, we add a constant energy production for every unit of mass, or as a function of radius $\epsilon(r) \sim \rho$.

We model convection by using the Ledoux stability criterion, with a mixing length $\alpha$ of 2.0, to determine whether convection occurs\cite{ledoux1947}. We also incorporate a small amount of exponential overshooting both above and below the convecting region. In addition to convection, we allow for additional transport through the star in the form of a stellar wind, modeled after \cite{brott2011}.

Once the initial main-sequence star with the appropriate mass is generated, it evolves under quasi-static evolution until a period of collapse occurs. When this collapse occurs, the evolution is treated using a Riemann equation solver, rather than the Newtonian solver used in the static evolution. This transition can occur when the star collapses in two slightly different ways, namely collapse due to fusion involving iron nuclei and collapse due to pair instability. While physically similar, the differences in conditions at which these types of collapse happen mean that the numerical treatment must be slightly different, and in particular, the triggers at which the transition from Newtonian to Riemann equation solving are different. 
To reflect the dynamics of core-collapse, the Riemann solver takes over when the temperature reaches $10^{9.6}$ K or neutrino luminosity reaches $10^{8.5}$~erg/g/s. In this case, neutrino luminosity is used as a proxy for the number of nuclear reactions that transform protons into neutrons, reactions that occur predominantly when high-mass nuclei around iron are involved. To capture pair-instability collapses, however, the Riemann solver is called if, evaluated only over the gravitationally bound portion of the star, the integral
\begin{equation}
	\int_0^{M_{max}} \left(\gamma-\frac{4}{3}\right) \frac{P}{\rho} dM
\end{equation} 
is less than zero. Here the function $\gamma = \partial\ln P / \partial\ln \rho$.

Because the range of phenomena that we wish to model includes pulsational pair instabilities, in which a star collapses due to the pair instability, but does not explode completely, we must allow for a period of hydrostatic evolution to commence after the conditions that necessitated hydrodynamic evolution are no longer fulfilled. At the same time, any material that is ejected should be removed in order to ease computational load. This process of removing the outer layers of the star occurs similarly to the process by which the initial main-sequence star was created. First, the outer layers of the star are dropped from the calculation, lowering the mass of the star. Layers are removed until the outermost gravitationally bound layer is reached. Then, the composition is adjusted to reflect any changes caused by resizing the discrete shells. Afterwards, the star is treated as before, evolving using the Newtonian solver as if this were a new star.

Finally, we employ two stopping criteria to detect the two possible outcomes that we consider. If the star is completely disrupted by a pair instability, the simulation will stop once the entire mass of the star is traveling outward faster than the local escape velocity. If, on the other hand, the star collapses in a core-collapse supernova, the simulation will stop once the iron core reaches a predetermined infall velocity, in our case $8 \times 10^8$~cm/s. Using this latter condition means that we do not follow the evolution of the star through the core collapse supernova, but rather stop during the collapse that leads up to it.

When computing the evolution of stars in this manner, we occasionally had to adjust various purely numerical parameters (identified within MESA as ``control parameters'') in order for the star to reach one of our acceptable end results. 
In general, our approach was to increase the number of spacial slices that were considered, and adjust various timing parameters so that a reasonable timestep was maintained throughout the evolution. In this way, we were able to evolve multiple stars with different amounts of extra energy, three of which we describe below.

\section{Comparison of Stars with and without Non-Nuclear Energy Component}\label{results}

\begin{figure}

	\includegraphics[width=\textwidth]{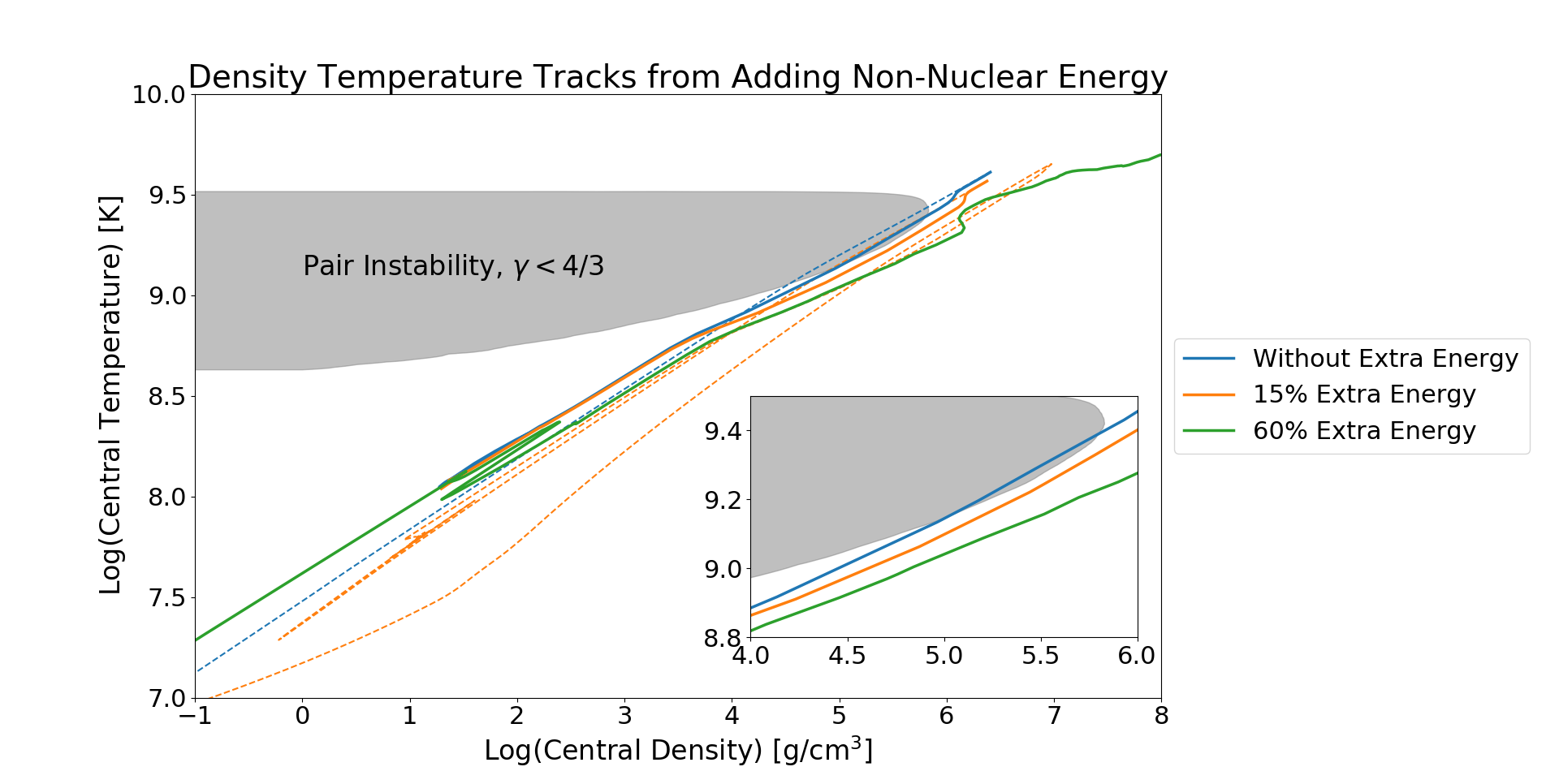}

\caption{
\label{trhoex}
Evolutionary behavior of $180 M_\odot$ stars when non-nuclear energy provides different fractions of the energy budget of the star. Like Figure~\ref{trho}, each track follows the evolution of the central density and central temperature initially from lower left to upper right. The blue curve follows the evolution of a star whose energy comes solely from nuclear reactions. This track is the same as the green $180 M_\odot$ track from Figure~\ref{trho}. When approximately 15\% of the energy produced by the star during the main sequence comes from non-nuclear sources, the star follows the orange evolution. After evolving through a pulsational pair instability supernova, the star completely destabilizes and leaves behind no remnant. When approximately 60\% of the energy produced during the main sequence is provided by non-nuclear sources, the star follows the green evolution. The feature on this (green) track occurring around a central density of $10^1$ to $10^2$~g/cm$^3$ occurs when the most interior layers of the stellar envelope mix into upper layers of the more evolved core below them. In this case of 60\% extra energy, the entire star misses the pair instability region, and instead evolves to a core collapse supernova and eventually a black hole remnant with a mass slightly below $119 M_\odot$.  
}
\end{figure}

\begin{figure}
	
	\includegraphics[width=\textwidth]{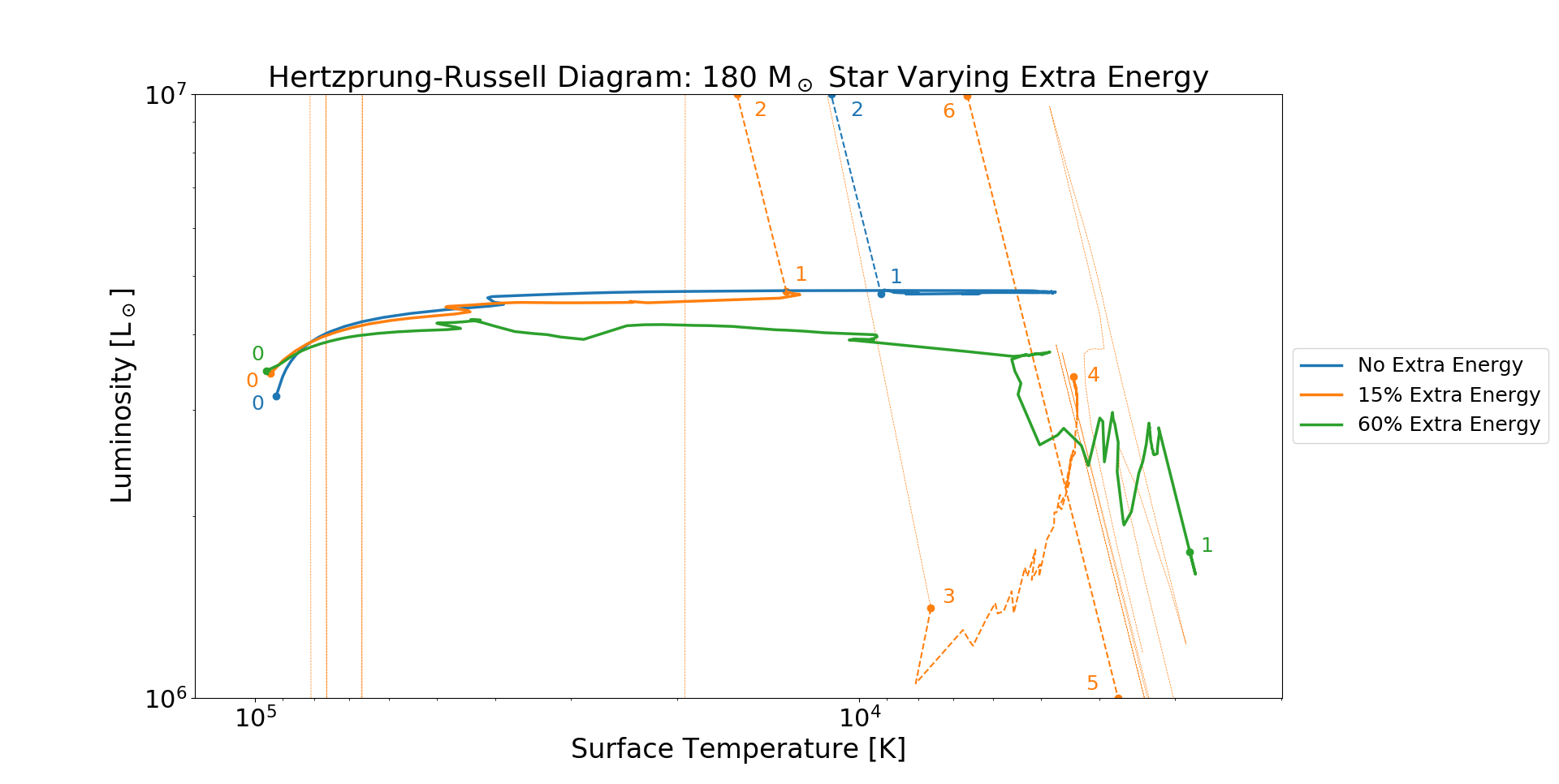}
	
\caption{
\label{hr}
Evolution through a Hertzprung-Russell diagram of $180 M_\odot$ stars with different amounts of non-nuclear energy. 
Each track starts on the main sequence on the left-hand side (labeled ``0'') and evolves initially to the right, as the star grows into a red giant phase. The blue track shows the evolution of a star solely powered by nuclear energy. The orange track shows the evolution of a star with 15\% of its energy from an extra non-nuclear energy source. And finally the green track shows the evolution of a star with 60\% of its energy from this non-nuclear source. Numbered points indicate the direction the star follows along each track. Dashed lines correspond to evolution after the (first) pair instability collapse (after the point labeled ``1'' on the blue and orange curves). The lighter dashed orange lines follow the evolution of the star with 15\% of its energy provided by non-nuclear sources while the star is undergoing periods of explosive mass ejection from the surface (between orange points ``2'' and ``3'' and between orange points ``4'' and ``5''), and may include the effect of numerical artifacts. In particular, the four vertical lines on the left side and center of the figure occur between ``2'' and ``3'' and correspond to the evolution after the pair instability explosion when the luminosity dropped to zero. The faint diagonal lines on the right side of the figure occur between ``4'' and ``5'' and correspond to a smaller mass ejection that happened independently of any pair instability effects. The portion of the orange dashed curve between points ``3'' and ``4'' corresponds to the evolution of the recollapsed remains of the star after it ejected some of its mass in the first pair instability explosion. Of particular note, while on the main sequence and shortly after leaving it, all three cases have similar observable properties. It is only later in the evolution of these stars that observable differences begin to arise.
}
\end{figure}

\begin{figure}
	
	\includegraphics[width=\textwidth]{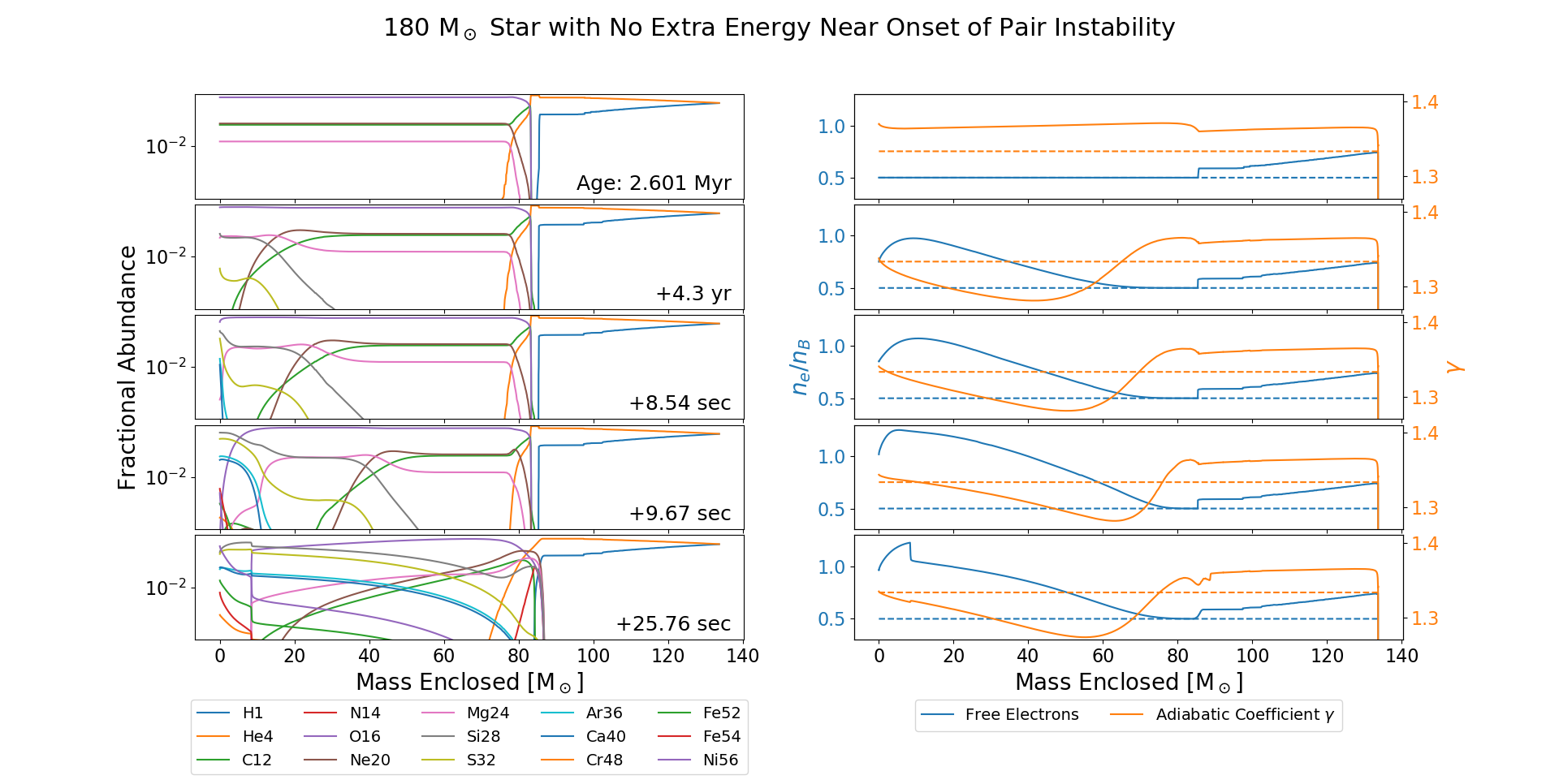}
	
\caption{
\label{pev0}
Evolution of a $180 M_\odot$ star powered only by standard nuclear energy at roughly the time of the pair instability. Left and right panels in each row show a snapshot of the conditions of the star at the same time. 
From top to bottom, each column shows the temporal evolution of the stellar interior from:
(top panel) moments before pair production becomes significant, (second panel from the top) shortly after pair production has caused the core to begin collapsing, (third panel from the top) as oxygen-burning first becomes non-negligible, (fourth panel from the top) approximately when the core stops collapsing, and (bottom panel) once the resulting shockwave has begun to expand just before the star begins to explode.
On the left, each panel shows the abundance of various isotopes as a function of mass enclosed within a sphere of given radius. 
On the right, each panel shows an adiabatic coefficient $\gamma$ in orange (values given on right vertical axis) and the free electron content (ratio of free electron to baryon number density) in the star in blue (values given in the left vertical axis). Instability occurs when $\gamma < 4/3$, marked by the orange dashed line, so any portion of the star with $\gamma$ below the dashed line is unstable to collapse. 
Any of the alpha-elements built up from fusing alpha particles should contribute $0.5$ free electrons per nucleus (demarcated by the dashed blue line) to the free electron content. Therefore, within the core of the star, pair production of electrons and positrons is occurring in any region where the free electron content is greater than the blue dashed line. The large region of the star which collapses due to the pair instability as well as the high number of free electrons in the core are characteristic of the evolution of a star that will ultimately undergo a pair instability supernova and completely explode, leaving no black hole remnant.
}	
\end{figure}

\begin{figure}
	
	\includegraphics[width=\textwidth]{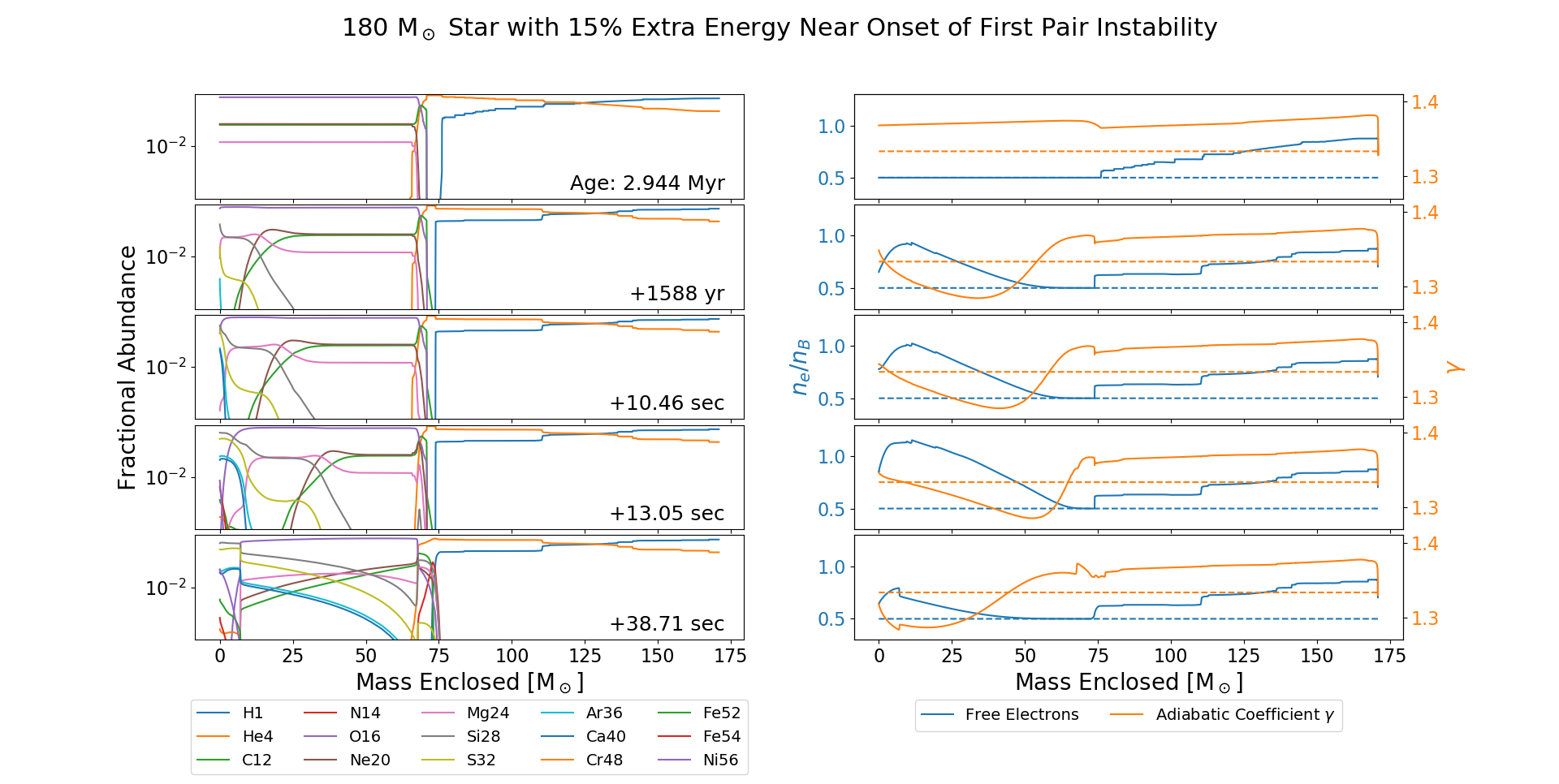}
	
\caption{
\label{pev08}
Evolution of a $180 M_\odot$ star with 15\% non-nuclear energy at roughly the time of the first pair instability. 
We use the same notation and a similar sequence of events as in Figure~\ref{pev0}, with time increasing from top to bottom panels.
As in Figure~\ref{pev0}, areas where $\gamma$ is below the orange dashed line are unstable to collapse, while areas where the free electron content are above the blue dashed line are experiencing electron-positron pair creation (at least within the core). Although still affecting a large portion of the core, the instability caused by production is less than that seen in Figure~\ref{pev0}. Likewise, the number of free electrons is slightly lower. Both are characteristic of a star that will ultimately undergo a pulsational pair instability supernova. In this case, that will lead to a total explosion of the star with no BH remnant.
}
\end{figure}

\begin{figure}
	
	\includegraphics[width=\textwidth]{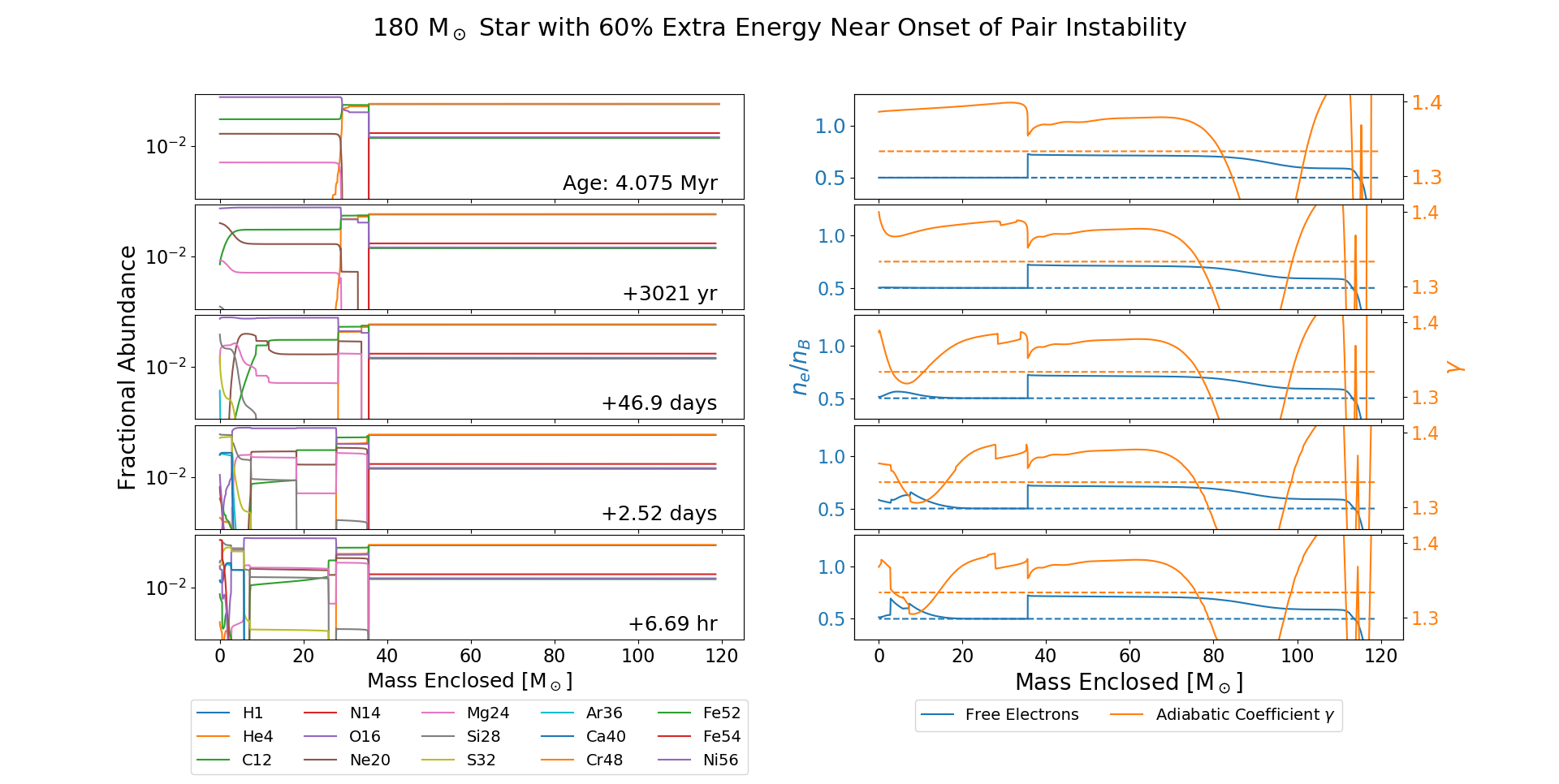}
	
\caption{
\label{pev22}
Evolution of a $180 M_\odot$ star with 60\% non-nuclear energy at roughly the time of the first pair instability. 
Again, the notation is the same as  in Figure~\ref{pev0}, and again time increases from top to bottom panels.
However, because the pair instability does not affect enough of the star to lead to a collapse of the core, we had to select the sequence of events differently. The top and bottom panels represent a time slightly before electron positron pair production (top) and after the star's interior has stabilized subsequent to  pair instability (bottom). The remaining panels show representative times between these two endpoints, but have little physical significance individually. 
On the right, $\gamma$ is plotted in orange and the free electron content in blue. The meaning of the dashed lines are the same as in Figures~\ref{pev0}. While there are large drops in the adiabatic coefficient in the stellar envelope, which likely occur where temperature is low enough to allow electrons to bind with nuclei, only a small region of the core enters the pair instability. Because this region is so small, the star remains stable throughout. Likewise, there is a region in which pair production occurs, but the number of electron-positron pairs produced remains small. Both of these phenomena are characteristic of a star that continues evolving to a core-collapse supernova and a black hole afterwards.
}
\end{figure}

For the purposes of this first study, we ran MESA stellar evolution simulations for the specific case of $180 M_\odot$ stars.  We simulated stars both with and without the addition of a new type of non-nuclear energy.
With only nuclear power providing support, as expected, we found that the  star undergoes a pair instability supernova, exploding in one burst and leaving no remnant. We then added different amounts of non-nuclear energy in the form of a constant $\epsilon_{DM}$ term and performed the simulations again.   Below, we compare the evolution seen in a few of these simulations.
Our simulations were for three cases: (i) the standard case of fusion power only for the star, as discussed in Section 5.1, (ii) a star with 15\% non-nuclear energy, in Section 5.2, and (iii) a star with 60\% non-nuclear energy in Section 5.3.
Our results will be illustrated in the following figures and tables:
 An overall description of the evolution can be illustrated by the central temperature-central density plot in Figure~\ref{trhoex}, which may be compared to Figure~\ref{trho}. In addition, internal properties of the stars in each of the simulations as they pass near the pair instability region can be seen in Figures~\ref{pev0}, \ref{pev08} and \ref{pev22}. The tracks that the stars take across a Hertzprung-Russell diagram in each simulation can be seen in Figure~\ref{hr}. Summaries of the evolutionary properties of the stars in each simulation may be found in Tables~\ref{sumtab} and \ref{evtab}.  

\begin{figure}
	
	\includegraphics[width=\textwidth]{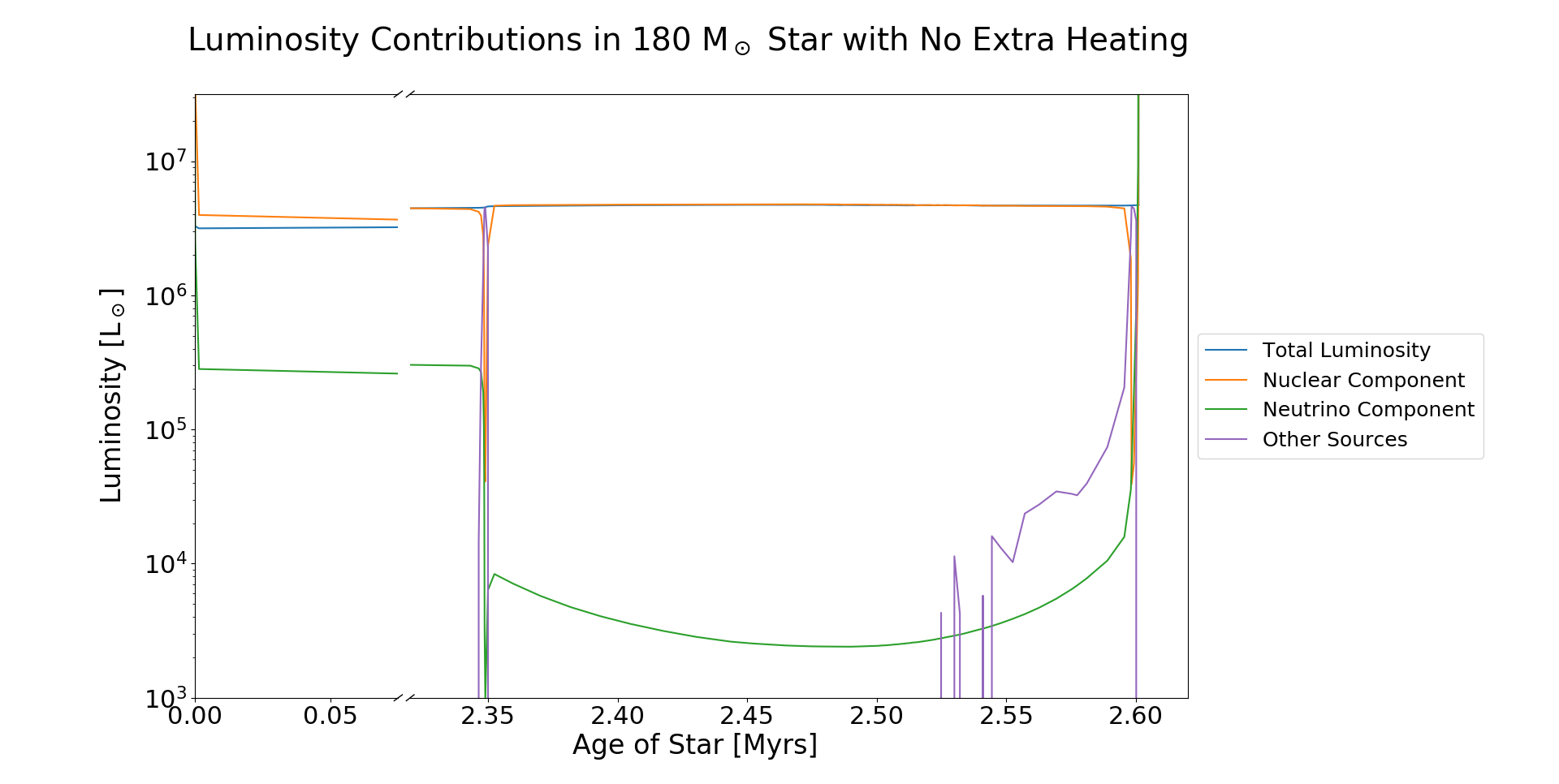}
	
\caption{
\label{lum0}
Breakdown of the luminosity of a $180 M_\odot$ star whose energy is provided solely from nuclear sources. The total luminosity output and three luminosity components are plotted as a function of the age of the star, focusing specifically on the early and late stages of the star's life. Much of the main sequence has been cut out, as the there is little change in any of the luminosity components during that period. The blue curve represents the total luminosity output by the entire star. The orange curve represents the luminosity attributed to all nuclear reactions, except for energy released in the form of neutrinos. The green curve represents specifically the luminosity output caused by neutrinos escaping the star. The purple curve represents all remaining sources of luminosity, which primarily comes from gravitational sources. For much of the star's evolution, nuclear energy dominates the energy in the star, so the two curves (blue and orange) become indistinguishable. The sharp drop in all components of luminosity around 2.36 Myr comes as the star finishes the main sequence and begins the helium-burning phase. As the star's evolution becomes less quasi-static, particularly near the end of its life, gravitational energy sources become more important. The sharp increase in luminosity at the end of the star's life is symptomatic of the explosive nature of the pair instability supernova.  Our simulations stop when all of the mass exceeds escape velocity from the star; in the end all the mass is ejected, leaving no remnant at all.
}
\end{figure}

\begin{figure}
	
	\includegraphics[width=\textwidth]{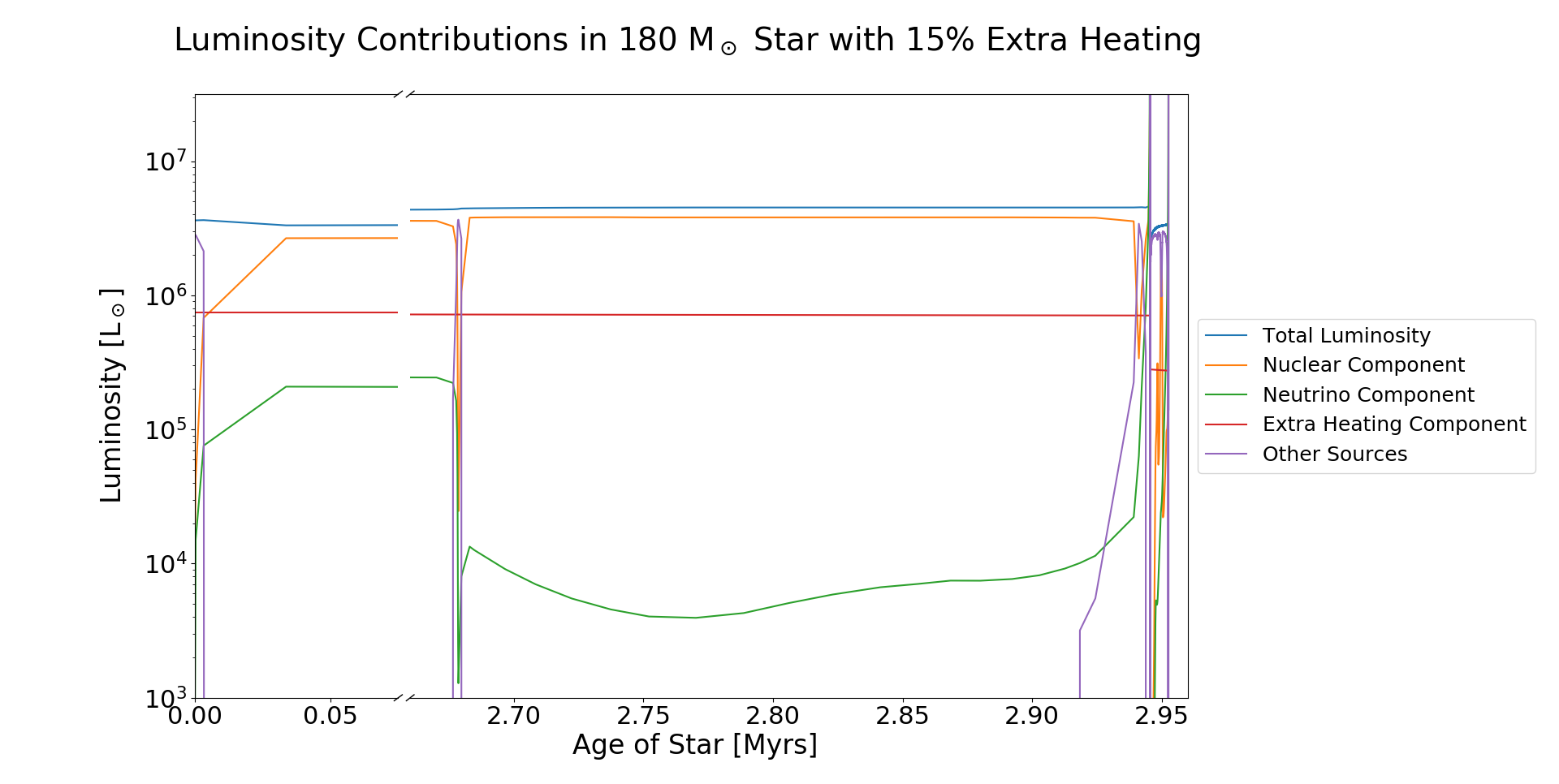}

\caption{
\label{lum08}
Breakdown of the luminosity of $180 M_\odot$ star with approximately 15\% of the main sequence energy due to non-nuclear sources. Like in Figure~\ref{lum0}, the total luminosity and its primary components are plotted as a function of time. In this case, the star in question has a constant non-nuclear energy source of $0.8 \times 10^4$~erg/g/sec throughout the star, which accounts for approximately 15\% of the energy produced during the main sequence. In this plot, much of the main sequence has been cut out, so as to focus on the more interesting evolution near the end of the star's life. The curves have the same meaning as in Figure~\ref{lum0}, with the addition of a red curve which follows the luminosity due to the non-nuclear energy source that has been added. The feature at approximately 2.68 Myr is caused by the star leaving the main sequence and beginning to burn helium. Near the end of the lifetime of the star there are two features, identifiable by the two sharp peaks in luminosity. These are each collapses and subsequent explosions associated with two separate pair instability events. Between those two events, much of the mass of the star has been lost, as is evidenced by the significant decrease in the component of luminosity due to the non-nuclear energy source, which is proportional to the mass of the star.  In the end all mass is ejected and there is no remnant left behind.
}
	
\end{figure}

\begin{figure}
	
	\includegraphics[width=\textwidth]{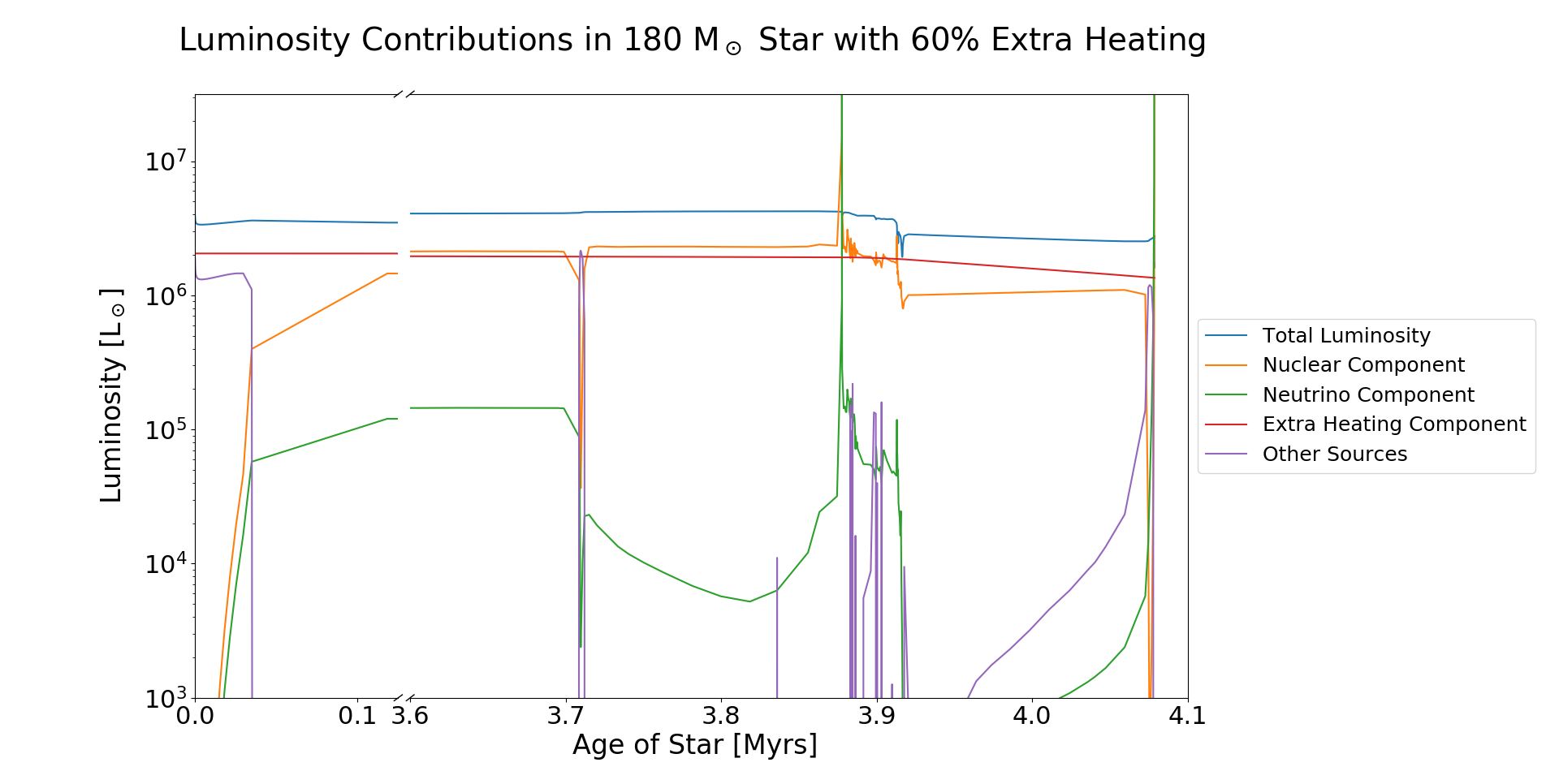}
	
\caption{
\label{lum22}
Breakdown of the luminosity of $180 M_\odot$ star with approximately 60\% of the main sequence energy due to non-nuclear sources. The various curves have the same meanings as in Figures~\ref{lum0} and \ref{lum08}. In this case, however, a constant energy source term of $2.2 \times 10^4$~erg/g/sec was added throughout the star. During the main sequence this accounts for approximately 60\% of the luminosity of the star.  Much of the main sequence was removed from this plot, focusing instead on less trivial eras during the beginning and end of the stellar lifetime. The feature at the beginning reflects a contraction onto the main sequence from the initial model generated by MESA. The feature occurring around 3.71 Myr is caused by the transition from burning hydrogen on the main sequence to burning helium afterward. The extended feature around 3.90 Myr is a result of changes to the stellar structure associated with a mixing of the inner stellar envelope into deeper layers of the star. This mixing caused a buildup of heavier elements in the outer layers of the star, enhancing the mass loss rate and leading to the gradual decline in the luminosity due to the extra non-nuclear component. The feature at the end of the lifetime of the star is associated with the collapse of the star into the beginnings of a core-collapse supernova.  The end result of this star will be a black hole remnant of mass slightly below $119 M_\odot$. 
}
\end{figure}
		
\begin{table}
	\begin{tabular}{|c|c|c|c|c|c|c|}
		\hline
		Percentage Non-Nuclear & Lifetime & \multicolumn{3}{c|}{Maximum Core Mass ($M_\odot$)} & Final mass & Supernova\\
		Energy Source & (Myr)  & He & C/O & Fe & ($M_\odot$) & Type\\
		\hline
		Nuclear-only & 2.601 & 86.53 & 82.07 & 0.0 & no remnant & PISN\\
		15\% & 2.953 & 77.02 & 68.22 & $5.9 \times 10^-3$ & no remnant & PPISN \\
		60\% & 4.079 & 56.21 & 28.91 & 2.01 & 119 & core collapse\\
		\hline
	\end{tabular}
\caption{
\label{sumtab}	
Summary properties of evolution in $180 M_\odot$ stars with and without added non-nuclear energy.  Lifetime corresponds to the total time between the initial main sequence model and the end of the simulation in either a complete explosion or the beginning of a core collapse supernova. In succession, the well-mixed central core of a star is predominantly composed of helium, carbon and oxygen, and iron. The maximal masses that each of these cores attain are listed. Final mass identifies the total gravitationally bound mass at the end of our simulations. A value of ``no remnant'' implies that at the end of the simulation, all of the mass of the star was unbound, meaning there would be no black hole remnant. Supernova type delineates the evolutionary track that the star undergoes, labeled by the type of supernova that it ends in, either pair instability (PISN), pulsational pair instability (PPISN), or core collapse. 
}

\end{table}

\begin{table}
	\resizebox{\textwidth}{!}{
	\begin{tabular}{|c||c|c|c|c|c|c||c|c|c|}
		\hline
		Percent. Non-nuc. & \multicolumn{6}{c||}{Burning stages} &  \multicolumn{3}{c|}{Pair Instability}\\
		\cline{2-7} \cline{8-10}
		Energy Source & H-burn & He-burn & C-burn & O-burn & Si-burn & Fe-burn & $\gamma< 4/3$ & Collapse begins & Collapse ends\\
		\hline
		Nuclear-only & 2.35 Myr & 0.25 Myr & 1800 yr & 7.2 sec & 40.9 sec & -- & $\mathrm{C} - 2.31$ days & +2.31 days & +49.9 sec \\
		15\% & 2.68 Myr & 0.26 Myr & 1600 yr & 5.3 sec & 60.2 sec & -- & $\mathrm{C} - 0.31$ days & +0.31 days & +37.6 sec \\
		60\% & 3.71 Myr & 0.37 Myr & 1700 yr & 1.2 days & 1.7 days & 2.8 hr & $\mathrm{C} - 5.5$ days & -- & -- \\
		\hline
	\end{tabular}
}
\caption{
\label{evtab}
Evolutionary timeline of $180 M_\odot$ stars with and without the addition of a non-nuclear energy source. The lifetime of the star is broken into six main nuclear burning stages, in which a particular reaction provides the dominant source of energy in the center of the star. H-burn is dominated by the pp- and cno-chains for fusing hydrogen into helium, He-burn by the triple-alpha process for fusing hydrogen into carbon, C-burn by the fusion of C-12 into Mg-24, O-burn by the reaction fusing O-16 into S-32, Si-burn in the dissociation of Si-28 into lighter elements, and Fe-burn in the dissociation of Fe-56 into lighter elements. While these are not the only fusion processes occurring, they are intended to give an estimate of the time scales involved. Where no time is given for the Fe-burn stage, this stage of burning was not reached, as there was an insufficient amount of iron produced in the core to allow iron-fusion to dominate in the core. In addition to the burning stages, information on pair instability is provided. $\gamma < 4/3$ corresponds to the time at which some portion of the star first experiences conditions in which $\gamma < 4/3$. This time is reported as a number of days prior to the end of the carbon-burning stage in all cases. The time at which (the first) pair instability collapse begins and is halted by a shockwave are provided as time differences from the $\gamma<4/3$ time and from when the collapse begins, respectively.
}
\end{table}

\subsection{Nuclear-only Star}

Using standard stellar evolution, with only fusion reactions as an energy source, we simulated the evolution of a $180 M_\odot$ star from main sequence to PISN. Hydrogen was fused in the core using both the pp chain and cno cycle until a helium core of mass approximately $86 M_\odot$ formed. In this core, hydrogen abundance was depleted to effectively 0, with He-4 becoming the dominant isotope until carbon began forming via the triple-alpha process. The main sequence of the star lasted approximately 2.35 million years while the helium burning phase lasted an additional 250 thousand years. The radius of the star during these periods ranged from $10^{12}$~cm to $10^{14}$~cm, with the majority of that radius being occupied by the envelope outside of the helium core. By the time helium-burning had ended, the star consisted of a $78 M_\odot$ carbon-oxygen core and a $55 M_\odot$ hydrogen-helium envelope. The remainder of the initial $180 M_\odot$ had been ejected from the star by stellar wind. 

Only hundreds of years after the star began burning carbon into magnesium and neon, the center of the star reached conditions necessary to begin producing electron-positron pairs. However, the effected area was initially small, and the star did not immediately begin collapsing. Within a year of pair production beginning though, the star had begun to collapse. Shortly after the outer parts of the core began to collapse, the innermost region began burning oxygen into silicon and sulfur. The energy produced by these reactions caused a shockwave to oppose the collapse only minutes after the nuclear reactions had begun. By the time this shockwave had begun, the carbon-oxygen core had reached $82 M_\odot$, while the total mass of the star had remained $134 M_\odot$.

Over the next several minutes, the star exploded. Although the radius at the surface did not change much, remaining approximately $10^{14}$~cm for the duration, the central density dropped from $10^{6.4}$~g/cm$^3$ to $10^{-3.4}$~g/cm$^3$. Presumably, this trend would have continued, but our simulation concluded here because all of the mass of the star had a velocity greater than the local escape velocity. This condition being satisfied implied that the star had undergone a PISN and would leave behind no remnant.

\subsection{Star with 15\% Non-Nuclear Energy}

By adding as little as~$0.8\times 10^4$~ergs/g/sec extra non-nuclear energy (approximately 15\% of the total energy produced by the star during the main sequence, as shown in the luminosity breakdown in figure \ref{lum08}), we begin to see noticeable differences in the stellar evolution, though the evolution still ends with no black hole remnant. The early lifetime of the star is similar to the case with only nuclear energy: hydrogen was fused leading to a helium core. However, with the non-nuclear energy, the helium core was only $77 M_\odot$. Furthermore, the main sequence lasted for 2.68 million years, slightly longer than the nuclear-energy-only case. The helium-burning phase added an additional 260 thousand years to this lifetime, before carbon-burning began. Interestingly, throughout this process, the radius ranged from only $10^{11}$~cm to $10^{13}$~cm, smaller than the nuclear-only case. By the time carbon-burning began, the carbon-oxygen core was $67 M_\odot$, and the hydrogen-helium envelope added an additional $104 M_\odot$. Clearly, by adding the non-nuclear energy, we moved much of the mass of the star from the core to the envelope.

Again, it takes only hundreds of years after the onset of carbon burning for the conditions in the core of the star to be appropriate for the production of electron-positron pairs. However, in the star with extra non-nuclear energy, the pair production began only after a non-negligible amount of oxygen-burning had already begun in the core. Like the case with only nuclear energy, it took only hours for the star to collapse after the center of the star began pair producing electrons and positrons. Curiously, while the core was collapsing, a small region within the envelope was expanding. This region may be associated with a hydrogen-burning shell that exists in a similar location.

Minutes later, a shockwave began that opposed the collapse. In the case including extra energy, the shockwave was due partially to oxygen-burning, and partially to silicon-burning. The silicon-burning occurred because the pair-instability was only reached after oxygen-burning had begun, meaning there was a non-negligible amount of silicon in the core. With no non-nuclear energy on the other hand, pair production happened before oxygen burning had begun, so there was no silicon-burning initially. By the time the shockwave had formed (in the star with non-nuclear energy), the carbon-oxygen core had reached $68 M_\odot$, while the total star mass has reached $170 M_\odot$.

Like the nuclear-only  case, the star then exploded. Over the next few hours, the star expanded, but in the case with a non-nuclear energy source, the entire mass of the star did not reach velocities greater than the local escape velocity. Instead, $103 M_\odot$ of material was carried off the star by the shockwave, leaving behind a $67 M_\odot$ gravitationally bound object. This object was partially supported initially by the decays of isotopes beyond iron, but eventually it contracted enough to once again fuse helium that had been produced by nuclear reactions involving silicon. As the star settled down, nuclei mixed via convection, causing small, temporary expansions and contractions that did not eliminate the general trend toward higher temperatures and densities.

Over a time scale of about ten thousand years, the cycle of nuclear processes drove this smaller star again towards the pair instability region. For this star, however, the deepest interior portion was an iron core, with nuclear fusion happening in a shell above this. By the time the star had begun to produce electron-positron pairs once again, it was only $66 M_\odot$, with a $47 M_\odot$ carbon-oxygen core. Although the entire star contained carbon and oxygen in various amounts by this point, the core was the mixed region in which the nuclear processes that involve carbon and oxygen take place.

Over a span of hours after the second onset of pair production, the star collapsed and exploded once again. This time, the explosion was caused predominantly by a burst of silicon-burning, though oxygen burning was a significant contributor to the energy imparted to the shockwave. From the onset of the shockwave, it took only hours for the star to completely destabilize and explode. Again, the entire star was moving faster than the local escape velocity and there was no remnant. By adding extra non-nuclear energy, we see the evolution of a star change from undergoing a PISN to a PPISN, but the end result remaining the same: no BH remnant survives.

\subsection{Star with 60\% Non-Nuclear Energy}

Adding $2.2\times 10^4$~erg/g/sec in extra non-nuclear energy (approximately 60\% of the total energy produced by the star during the main sequence, as seen in Figure~\ref{lum22}), we saw a $180 M_\odot$ star begin to collapse toward a core-collapse supernova, suggesting that it would become a black hole within the mass gap.
As with the previous two cases, the early life of the star was fairly unremarkable. Hydrogen fused into helium in much the same way, but with the addition of a larger amount of energy, the lifetime on the mean sequence was further extended. For this case, the main sequence lifetime was 3.71 million years, and left a helium core of only $56 M_\odot$. The helium-burning phase began similar to the other cases, but approximately 100 thousand years after it began, there was a phase that was not seen in the other cases. In this phase, material in the hydrogen-rich envelope was drawn down into the outer layers of the helium-carbon-oxygen core. As a result, intense hydrogen burning provided a strong source of energy for the star. Because of the burst of energy, the star expanded and the core cooled. This mixing also allowed carbon and oxygen from the core to migrate out to the envelope, where it was processed by the CNO cycle into a mix of carbon, nitrogen, and oxygen. This mix of chemicals might provide a visible signal to hint at non-standard evolution, if a stellar spectrum is taken later in the star's life. In addition, because of the mixing of the carbon and oxygen out of the core, the helium-carbon-oxygen core ultimately decreased in mass, reaching approximately $36 M_\odot$ before the process was complete. Overall, this process took approximately 50 thousand years, after which the star returned to the regular pattern of nuclear processes occurring in the core.

By the time carbon-burning began, the life of the star was 4.08 million years, but the star had decreased in total mass from $180 M_\odot$ to $119 M_\odot$. The carbon-oxygen core, meanwhile, had reached only $28 M_\odot$. It then took only months for oxygen-burning to begin. Days after the onset of oxygen burning, portions of the core of the star did experience the conditions necessary for electron-positron pairs to be formed, but too little of the star was affected to cause any large-scale instabilities. Curiously, there was a small shock that began near the boundary between the envelope and helium core. It persisted until the end of the simulation, but was always small compared to the escape velocity.

In any case, as the star continued to evolve over the next several hours or days, the core became more iron-rich, and the region in which fusion occurred was pushed outward into shells around this iron core. The temperature of the star increased, as did its density until the iron in the core became involved in nuclear reactions. Within a second, the iron core in the star began to collapse, leading to the onset of a core-collapse supernova. When our simulations ended, the mass of the star was $119 M_\odot$, and none of it was moving outward with a velocity greater than the local escape velocity. As is typical in core-collapse supernovae, it is likely that some of this mass will be lost before a black hole is formed, but at least some portion of it will likely survive to form a black hole.

\section{Conclusion}

According to standard stellar evolution models, where
a star's only power source is nuclear fusion, it is impossible for black holes in a mass range $50 M_\odot$ to $140 M_\odot$ to form. Stars with masses that could lead to black holes in that range ($150 M_\odot$ to $240 M_\odot$) instead become unstable as a result of electron-positron pair production, ultimately exploding entirely, leaving behind no black hole remnant.
However, the gravitational wave event GW190521 recently observed by the LIGO and Virgo collaborations led to the discovery of black holes with masses $66 M_\odot$ and $85 M_\odot$, which lie within the pair instability mass gap.  
In standard stellar evolution, the lack of black holes in this mass range seems to be tied to the temperature- and density-dependence of nuclear fusion, which can take place only near the hot dense centers of stars.

In this paper we introduced an additional energy source of non-nuclear origin, which produces energy equally throughout the star.
With this new energy source, it is possible to circumvent the pair instability entirely, allowing stars to evolve into  black holes inside the mass gap. 
As a concrete example, we used the MESA stellar evolution code to study  $180 M_\odot$ stars with and without an extra non-nuclear power source.
In standard stellar evolution, a star of this mass undergoes a pair instability supernova at the end of its life, leaving no black hole remnant. However, 
we found that
introducing a source of energy that is independent of temperature and density can cause a star of this mass to evolve differently. 
When as little as 15\% of the energy produced in the star (measured when the star is on the main sequence) is provided by non-nuclear energy, noticeable evolutionary changes occur;
yet the end result is still a complete explosion leaving no black hole remnant.  Instead of being completely destabilized in one explosion, the star undergoes multiple pulses, each of which eject mass until the star is completely destroyed. 
When approximately 60\% of the energy produced in the star (on the main sequence) is from a non-nuclear source, the star evolves to avoid the pair-instability-induced explosion and instead evolves to undergo a core-collapse supernova that leaves behind a black hole remnant. In the case of a $180 M_\odot$ star with 60\% of the energy coming from a non-nuclear source, the supernova precursor mass is approximately $119 M_\odot$, implying that a black hole remnant would be slightly smaller, but still comfortably within the mass gap. 
From this example, we see that introducing non-nuclear energy sources to stars can prevent their destruction because of pair instability,  can fill in the BH ``mass gap,'' and can thereby explain
the 66 and $85 M_\odot$ black hole masses observed by LIGO and Virgo in GW190521.  

Our studies identified the localization of energy production in the star as a key ingredient in whether or not a pair instability SN destroys the star. 
In the case of nuclear fusion only, the power source is centrally concentrated 
in the stellar core due to the requirement of high temperature for fusion; in this case a SN explosion completely destabilizes the star, leaving no remnant.
With the addition of non-nuclear energy that is more evenly spread out throughout the star, on the other hand, we find that a BH remnant survives.
In particular, as more non-nuclear energy is added to the star, the mass in the core decreases while the mass in the envelope increases. Changing the mass distribution like this is consistent with the evolutionary changes we see, since evolutionary behavior depends on core mass, while the mass of any remnant left behind depends on the star's total mass. In order for an injection of energy to lead to this sort of shift in mass distribution, the energy source cannot respond the same way to temperature and density as nuclear reactions. We explored the evolutionary effects of introducing an energy source that was independent of density and temperature, but it is likely that energy sources with weaker dependence on temperature and density can lead to similar, albeit weaker effects. In a similar way, while we used a model energy source which is constant in time, it may be possible for energy sources that vary in time to cause similar evolutionary changes, but only if the energy source alters the temperature and density properties within the star as the star nears the pair instability. In principle, the range of properties that energy sources may take and still cause changes to the evolution suggests that introduction of any of numerous mechanisms may allow black holes to form in the mass gap. 

While we remain agnostic as to the mechanism responsible for the non-nuclear energy source in the star, a
 concrete example is the annihilation of dark matter (e.g. WIMPs or some SIDM particles).
Previous work \cite{spolyar2009, freese2016} has shown 
that Dark Stars with masses in the range $1-10^7 M_\odot$ can be fully supported by dark matter annihilation heating. Thus, DM annihilation is capable of providing the amount of energy required of the non-nuclear heat source we introduce.
Furthermore, the energy source from DM annihilation is independent of temperature and significantly less dependent on density than nuclear reactions. Taken together, these two properties suggest that dark matter annihilations would pose a good candidate for the energy source that would allow stars to evolve into black holes in the mass gap. 
Further, it is interesting to speculate that it may be possible to deduce from observations of black holes properties of their precursor stars, and from these stars say something about dark matter physics.  

In summary, we find that the addition of non-nuclear energy sources in stars may provide an explanation of black holes in the mass gap, such as  those recently discovered by the LIGO and Virgo Collaborations.

\begin{acknowledgments}
	We thank Chris Kelso and Pearl Sandick for useful discussions at the initiation of this project. We also thank Michael Montgomery and Tanja Rindler-Daller for assistance getting started using MESA.
	KF is the Jeff and Gail Kodosky Endowed Chair in Physics at the University of Texas, Austin; KF and JZ thank the Chair funds for support.  KF acknowledges support by the Vetenskapsr\r{a}det (Swedish Research Council) through contract No. 638-2013-8993 and the Oskar Klein Centre for Cosmoparticle Physics. KF acknowledges support from the Department of Energy through DoE grant DE-SC0007859 and the Leinweber Center for Theoretical Physics at the University of Michigan.  KF would like to thank Perimeter Institute for hospitality (KF is supported by the Distinguished Visitors Research Chair Program).
\end{acknowledgments}

\bibliography{PartialDarkStarPaper}

\end{document}